\documentclass[prl,twocolumn,showpacs,amsmath,amssymb]{revtex4-1}	
\usepackage{graphicx}
\usepackage{dcolumn} 
\usepackage{bm}
\usepackage{epsfig}
\usepackage{braket}
\usepackage{hyperref}
\hypersetup{
    colorlinks=true,        
    linkcolor=blue,          
    citecolor=blue,        
    filecolor=magenta,      
    urlcolor=blue           
}

\def\laco*{LaCoO$_3$}
\def\ehs*{$\epsilon_{\text{HS}}$}
\def\eis*{$\epsilon_{\text{IS}}$}
\def\els*{$\epsilon_{\text{LS}}$}

\newcommand{\bk}{\mathbf{k}}
\newcommand{\citenamefont}{}
\newcommand{\bibnamefont}{}
\def\s*{$|s\rangle$}
\def\t*{$|t_\alpha\rangle$}
\def\se*{$s$}
\def\te*{$t_{\alpha}$}
\def\jo*{$j_{1/2}$}
\def\jt*{$j_{3/2}$}

\usepackage{color} 				
\usepackage[normalem]{ulem} 			

\begin{document}
\title{Field-induced exciton condensation in LaCoO$_3$}

\author{A. Sotnikov}
\email{sotnikov@fzu.cz}
\affiliation{Institute of Physics,
Academy of Sciences of the Czech Republic, Na Slovance 2,
182 21 Praha 8, Czech Republic}
\author{J. Kune\v{s}}
\affiliation{Institute of Physics,
Academy of Sciences of the Czech Republic, Na Slovance 2,
182 21 Praha 8, Czech Republic}

\date{\today}

\begin{abstract}
Motivated by recent observation of magnetic field induced transition in LaCoO$_3$ we study the effect of external
field in systems close to instabilities towards spin-state ordering and exciton condensation. We show that, 
while in both cases the transition can be induced by an external field, temperature dependencies of the critical 
field have opposite slopes. Based on this result we argue that the experimental observations select
the exciton condensation scenario. We show that such condensation is possible due to high mobility
of the intermediate spin excitations. The estimated width of the corresponding dispersion is large enough
to overrule the order of atomic multiplets and to make the intermediate spin excitation propagating 
with a specific wave vector the lowest excitation of the system.
\end{abstract}
\maketitle

Perovskite cobalt oxide \laco* exhibits unusual physical properties that have attracted attention
for over half a century. Fine balance between crystal field splitting and ferromagnetic Hund's
coupling places the Co$^{3+}$ ions to the vicinity of a spin-state transition. In combination with rather
covalent Co-O bonds and electron-lattice coupling it makes \laco* a complicated physical system.  
The electrical conductivity and magnetic susceptibility divide the $T$-dependent phase diagram
of \laco* into three regions with crossovers in-between: diamagnetic insulator ($<80$~K), paramagnetic insulator,
and paramagnetic metal ($>600$~K). Several approaches have traditionally been used to describe
the physics of \laco*: (i)~the single-ion picture of low spin ($S=0$, LS) ground state of the Co$^{3+}$ ion
with the  intermediate spin ($S=1$, IS) or high spin ($S=2$, HS) excitations~\cite{zobel02,knizek09} augmented with
spin-exchange between these states on the lattice~\cite{yamaguchi96}, (ii)~band structure approaches with electrons
interacting via static mean field~\cite{abbate94,korotin96,knizek05} and (iii)~combination of both in the form of dynamical mean-field
theory~\cite{eder10,zhang12,krapek12}. Despite numerous studies, underlying physics of \laco* remains an open problem.

The intermediate temperature regime with a paramagnetic susceptibility, a charge gap and no sizeable Co-O bond-length disproportionation
is particularly difficult to describe. Existing theories either yield a metallic state~\cite{korotin96} or exhibit a spin-state order (SSO)~\cite{bari72,knizek09,kunes11,karolak15},
a periodic arrangement of Co atoms in different spin states  necessarily accompanied by Co-O bond-length disproportionation. 
Recent high magnetic fields experiments~\cite{altarawneh12,rotter14,ikeda15}, which found metamagnetic transition above 50~T, 
provide an important clue as to the nature of the intermediate temperature regime.  
As pointed out by authors of Ref.~\onlinecite{ikeda15} the increase of the critical field $h_c$
with temperature $T$ is counterintuitive given the fact that increasing temperature promotes population of the magnetic states.

\begin{figure}
\includegraphics[width=\linewidth]{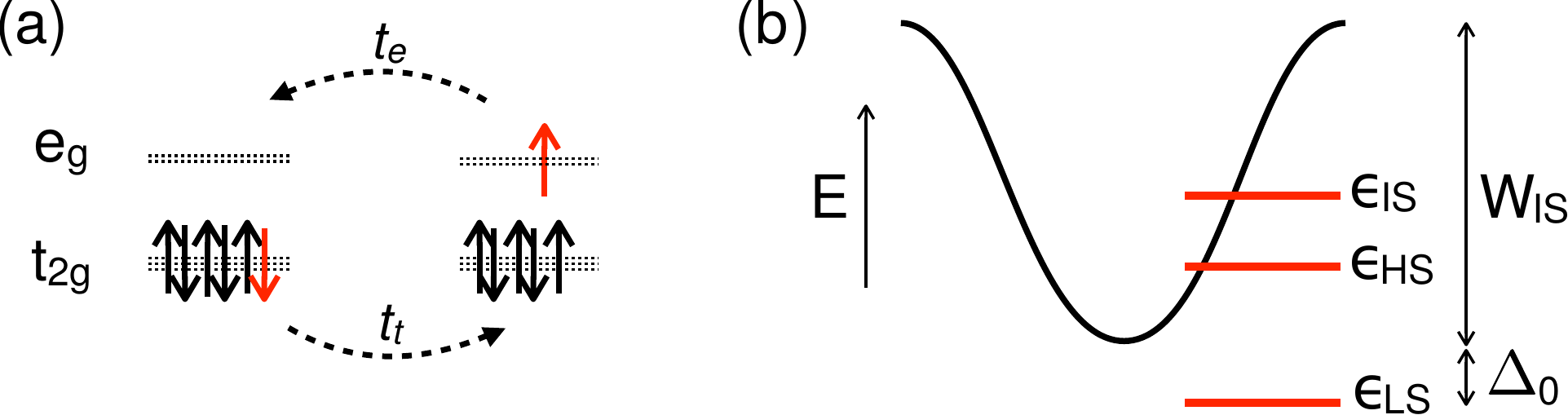}
\caption{\label{fig:cartoon} (a) Nearest-neighbor hopping process that gives rise to IS-LS exchange. (b) Cartoon of the atomic multiplet energies
together with the dispersion of a single IS state on the LS background.}
\end{figure}

In this paper we show that mobility of IS excitations on the LS background, an aspect missing 
in existing theories, plays an important role in \laco*. Estimates based on first principles calculations show that while 
the HS excitations are essentially immobile, the bandwidth of the IS dispersion is of the order of several 100~meV. 
The lowest excitation in solid, therefore, may have different character than the lowest single-ion excitation
as depicted in Fig.~\ref{fig:cartoon}. The nature of the low-lying excitations can be probed by the field-induced transition. 
Immobile excitations favor formation of SSO. Mobile excitations, on the other hand, lead to formation of a homogeneous 
excitonic condensate (EC). While both SSO and EC can be induced by magnetic field, we show that their $h_c(T)$ dependencies 
have opposite slopes.

In the following, we use dynamical mean-field theory (DMFT) to study the minimal model allowing for a spin-state transition~\cite{kunes14a}.
We show that both SSO and EC can be induced by an external magnetic field and calculate the temperature dependencies of the
critical field  $h_c(T)$ together with other relevant physical observables close to the phase boundaries. 
To demonstrate the feasibility of the EC scenario, we use density functional band structure analysis to estimate the dispersion of the IS and HS excitations in real material.

{\bf Model.} 
We study the two-orbital Hubbard model on a square lattice, which exhibits both the instability
towards formation of SSO and EC~\cite{kunes14a}. With an external magnetic field~$h$ the Hamiltonian reads
\begin{eqnarray}
 {\cal H} &=& \sum_{\alpha\beta}t_{\alpha\beta}\sum_{\braket{ij}\sigma} (c^\dag_{i\alpha\sigma}c_{j\beta\sigma} + \text{H.c.})
 \nonumber\\
 && 
 + \sum_{i}{\cal H}^{(i)}_{\text{int}}
 + \sum_{i\alpha\sigma}(h\sigma-\mu_\alpha)n_{i\alpha\sigma},\label{H}
\end{eqnarray}
where $c^\dag_{i\alpha\sigma}$ ($c_{i\alpha\sigma}$) are the fermionic creation (annihilation) operators acting at the lattice site $i$, $\alpha=\{a,b\}$ represents the orbital index, $\sigma=\{1/2,-1/2\}$ denotes the electron spin projection on the magnetic-field axis (in units of $\hbar=1$), the $2\times2$ symmetric matrix $t_{\alpha\beta}$ consists of the amplitudes for the usual (intra-orbital, $\alpha=\beta$) and the cross (inter-orbital, $\alpha\neq\beta$) hopping processes, and the notation $\braket{ij}$ indicates the summation only over nearest-neighbor sites. The local interaction part~${\cal H}^{(i)}_{\text{int}}$ is chosen to have only density-density contributions,
\begin{eqnarray}
 {\cal H}^{(i)}_{\text{int}} &=& U\sum_{\alpha} n_{i\alpha\uparrow}n_{i\alpha\downarrow} + (U-2J)\sum_{\sigma} n_{ia\sigma}n_{ib-\sigma}
 \nonumber\\
 &&
 +(U-3J)\sum_{\sigma} n_{ia\sigma}n_{ib\sigma},
\end{eqnarray}
and in the last term $\mu_{a,b}=\mu\pm\Delta/2$, where $\mu$ is fixed to yield average filling of two electrons per lattice site and $\Delta$ is referred to the crystal-field splitting.

We use dynamical mean-field theory (DMFT) \cite{Geo1996RMP} with the continuous-time quantum Monte Carlo hybridization-expansion (CT-HYB) impurity solver \cite{Wer2006PRL,Gul2011RMP} in the so-called segment representation modified to include off-diagonal hybridization important for an account of the excitonic instability close to the spin-state crossover \cite{kunes14a,kunes14c}.
This approach allows to obtain the Green's functions of the impurity problem, therefore, gives an access to spectral characteristics and corresponding local observables, e.g., the correlators $\braket{c^\dag_{i\alpha\sigma}c^{\phantom\dagger}_{i\beta\sigma'}}$ that provide with information on the magnetization (by taking the elements with $\alpha=\beta$) and the EC order parameters ($\alpha\neq\beta$). In the homogeneous case (single unit cell) we focus on the EC order parameters of the type $\phi^{+}=\braket{c^\dag_{ia\uparrow}c^{\phantom\dagger}_{ib\downarrow}}$ and $\phi^{-}=\braket{c^\dag_{ia\downarrow}c^{\phantom\dagger}_{ib\uparrow}}$ and the magnetization $M=\sum_\alpha\sum_\sigma\sigma n_{i\alpha\sigma}$ with $n_{i\alpha\sigma}=\braket{c^\dag_{i\alpha\sigma}c^{\phantom\dagger}_{i\alpha\sigma}}$.
In case of SSO, we also study the difference in occupations of the $a$-orbital on two neighboring sites $i$ and $j$, $d_a=\sum_\sigma|n_{ia\sigma}-n_{ja\sigma}|$.

\begin{figure}
\includegraphics[width=\linewidth]{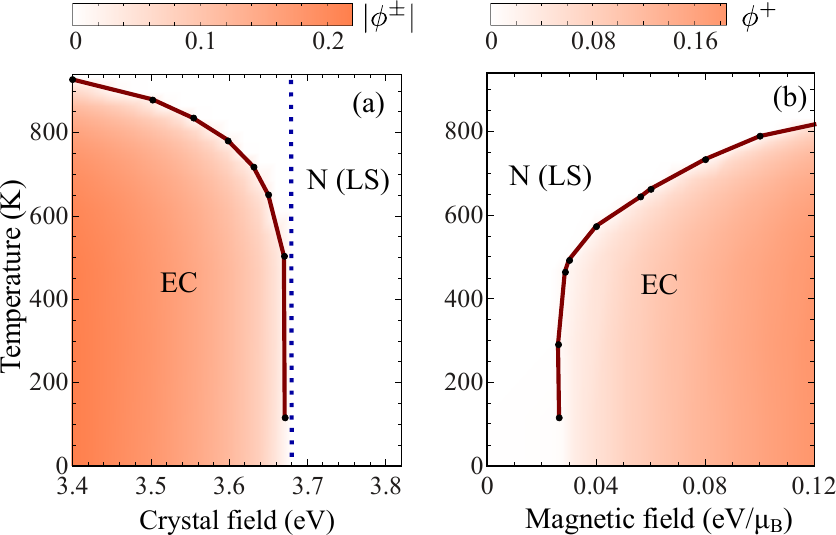}
    \caption{\label{fig2}
    Dependence of the critical temperature $T_c$ on the crystal-field splitting at $h=0$ (a) and on the magnetic field~$h$ at $\Delta=3.68$ (b). Other parameters are $U=4$, $J=1$, $t_{aa}=0.4118$, $t_{bb}=-0.1882$, and $t_{ab}=0.05$.}
\end{figure}

First, we analyze the phase diagram of EC in the external magnetic field. We start from parameters of Ref.~\onlinecite{kunes14c} where the excitonic
condensation was found around 850~K. We increase the crystal field so that the transition is suppressed, see Fig.~\ref{fig2}a, and 
choose the crystal field $\Delta$ so that the system is in the normal [N(LS)] phase, but close to the phase boundary. Then, we apply an external magnetic field $h$.
Upon reaching a critical field $h_c$ the system undergoes a transition to the ferromagnetic excitonic condensate characterized by non-zero value of $|\phi^+|>|\phi^-|$ for positive $h$.
The corresponding magnetization curves that are shown below in Fig.~\ref{fig:entropy}a exhibit a linear increase with $h$ at low fields and have a slope (uniform susceptibility) increasing with temperature. The excitonic condensation
is reflected by a kink at $h_c(T)$ in the $M(h,T)$ curves at constant temperature. The complete $h$-$T$ phase diagram in Fig.~\ref{fig2}b shows a clear increase of $h_c$ with increasing temperature.

The behavior of the present model is easily understood by considering the strong coupling limit. At $T=0$ and $h=0$ the excitation spectrum is described by the exciton band with a finite gap (see Fig.~\ref{fig:cartoon}).
At nonzero field $h$, the exciton band experiences the Zeeman splitting, thus as soon as the gap closes, the condensate starts to develop. 
At nonzero $T$, a larger $h_c$ is required to form EC in order to overcome the higher entropy of the normal phase,see Fig.~\ref{fig:entropy}. The entropy differences are calculated accordingly to the thermodynamic relations $\Delta S(h) = \int_{0}^{h} \left(\frac{\partial M}{\partial T}\right)_{h'} dh'$ and $\Delta S(T) = \frac{E(T)}{T}-\int_{T_0}^{T}\frac{E(T')}{T'^2} dT'$.
The increase of the uniform susceptibility in the normal phase with temperature reflects the single-ion physics of thermal population of the spinful states.
We point out that even in the presence of an external magnetic field the excitonic condensation is a phase transition in the thermodynamic sense 
as it breaks the spin-rotational symmetry around the direction of the external field ($z$-axis in the present model). 
\begin{figure}
\includegraphics[width=\linewidth]{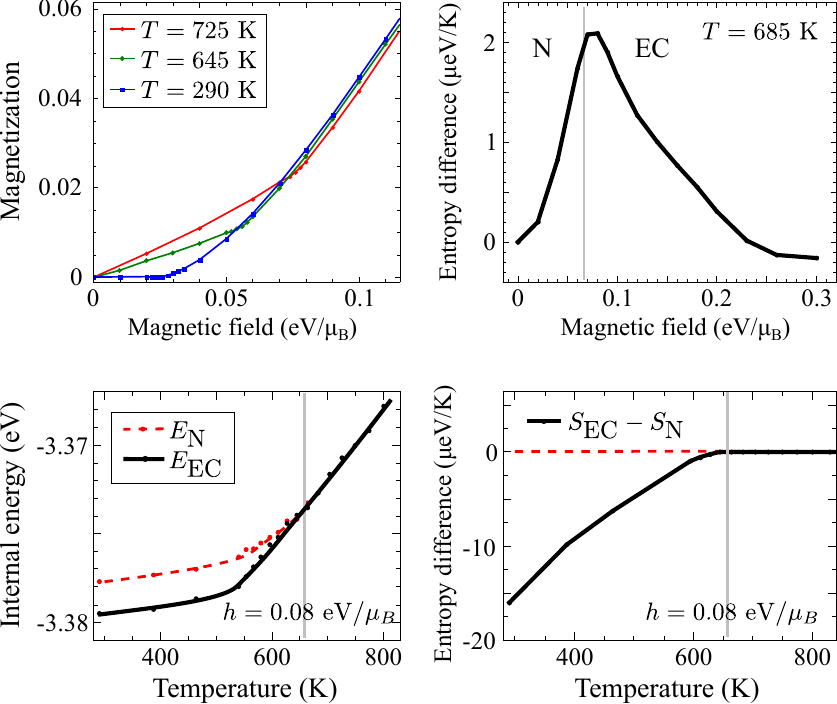}
    \caption{\label{fig:entropy}
    Dependencies of the magnetization and the entropy difference on the strength of the external magnetic field (upper row) and dependencies of
    the internal energy $E$ and the entropy difference on the temperature at constant magnetic field (lower row). Other parameters are taken the same as in Fig.~\ref{fig2}.}
\end{figure}

\begin{figure}
\includegraphics[width=\linewidth]{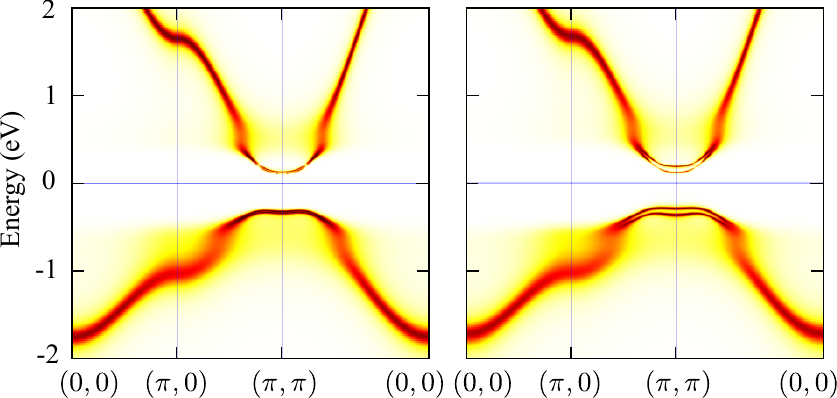}
    \caption{\label{fig:spectra}
    One-particle spectral densities $A(\bk,\omega)$ calculated with the maximum-entropy method \cite{Jarrell1996PR} in the normal (left) and EC (right) phases of Fig.~\ref{fig2}(b) at $T=290$~K, $h=0.02$~eV/$\mu_{\text{B}}$ and $h=0.05$~eV/$\mu_{\text{B}}$, respectively.}
\end{figure}

In order to demonstrate that the field-induced transition is not connected to any significant changes or closing of the one-particle gap we have calculated the 
one-particle spectra above and below $h_c$. The spectra in Fig.~\ref{fig:spectra} show that the Zeeman splitting of the electronic bands
is order of magnitude smaller than the charge gap.  

\begin{figure}
\includegraphics[width=\linewidth]{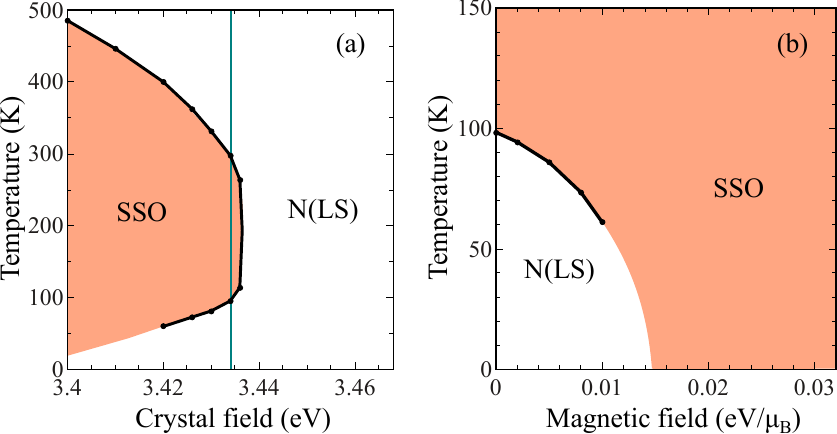}
    \caption{\label{fig:hsls}
    Dependence of the transition temperature between LS homogeneous and HS-LS disproportionated (SSO) phases on the crystal-field splitting at $h=0$ (a) and on the magnetic field~$h$ at $\Delta=3.434$ (b). Other parameters are $U=4$, $J=1$, $t_{aa}=0.45$, $t_{bb}=0.05$, and $t_{ab}=0$.}
\end{figure}

Next, we discuss the alternative SSO scenario.  
To this end we chose the asymmetric hopping parameters of Ref.~\onlinecite{kunes11} 
and a crystal field $\Delta$ close to the ``tip of the belly'' in Fig.~\ref{fig:hsls}a,
for which the SSO transition is reentrant and the exciton condensation is suppressed~\cite{kunes15}. 
The strong coupling limit provides a simple understanding of the reentrant behavior~\cite{kunes11}. Starting from the purely LS state at $T=0$, atoms in the HS state are generated randomly with increasing temperature. At the lower $T_{c}$ the concentration of HS sites becomes so high that the HS-HS repulsion~\cite{kunes15} drives the system into the ordered state that eventually melts at the upper $T_{c}$. In an external magnetic field~$h$, the solid phase expands. 
This is particularly pronounced for the low-$T$ transition from the normal to the solid phase, for which $T_{c}$ becomes more suppressed with increasing field~$h$ as shown in Fig.~\ref{fig:hsls}b. The sign $dh_c/dT$ is related to the fact that the transition from the low-$T$ (normal) to the high-$T$ (solid) phase is driven by the internal energy but not the entropy.

The temperature scales of the SSO and EC transitions are controlled by the values of parameters $t_a^2+t_b^2$ and $t_at_b$, respectively~\cite{kunes14a}. Varying the ratio $t_a/t_b$ one can
therefore manipulate the extent of the SSO and EC phases~\cite{kunes15,tatsuno16}. While the present model captures the essential physics of the field-induced transition,
it is too simplified to provide quantitative estimates of transition temperatures for the real material. In particular, the orbital degeneracy in the real material and non-local
fluctuations, which are absent in DMFT treatment, reduce the EC transition temperature.

{\bf Material.} 
The main difference of real \laco* from the studied model is the orbital degeneracy, which gives rise to excited states with both $S=1$ (IS) and $S=2$ (HS).
At low temperatures, when the concentration of excitations is low, the key difference between the IS and HS excitations is their mobility on the LS background. First, we estimate the 
amplitude for propagation of the IS excitations given by the second-order hopping process shown in Fig.~\ref{fig:cartoon}. We consider IS excitations with the $T_{1g}$ orbital symmetry, which have lower excitation energy and higher mobility than their $T_{2g}$ counterparts \cite{Dresselhaus2008,kunes14b}. The $T_{1g}$ excitation can be viewed as a bound pair of the butterfly-shaped $t_{2g}$ hole and $e_g$ electron rotated by 45$^\circ$,
i.e., $xy\otimes x^2-y^2$,  $yz\otimes y^2-z^2$ and $zx\otimes z^2-x^2$. This geometry makes the $T_{1g}$ excitations mobile in the plane of the orbitals with practically no hopping in the perpendicular direction. The in-plane nearest neighbor hopping amplitude can be estimated as
\begin{equation}
t_{\text{IS-LS}}=\frac{2 t_t t_e}{\tilde{U}}.
\end{equation}
Here, $t_t=-0.45$~eV  and $t_e=-0.16$~eV are the electronic $xy\leftrightarrow xy$ and $x^2-y^2\leftrightarrow x^2-y^2$ hopping amplitudes, respectively, in the $x$ or $y$ directions. The numerical values
were obtained from $d$-only Wannier functions calculated in the idealized cubic structure~\cite{wien2k,wien2wannier, wannier90}.  
The parameter $\tilde{U}$ is approximately $U-2J$ and we estimate its value for $d$-only model to be in the range of 2-4~eV. This leaves us with 
$t_{\text{IS-LS}}$ in the range of 36--72~meV and the bandwidth of IS dispersion indicated in Fig.~\ref{fig:cartoon}b (considering four in-plane nearest neighbors) in the range of 290--580 meV.
Absence of the condensed state for $h=0$ implies a finite excitation gap and puts the energy of atomic IS excitation to at least half the bandwidth, i.e., $4t_{\text{IS-LS}}$.
This estimate is consistent with the experimental measurements~\cite{haverkort06} and allows to place the energy of atomic HS excitation 100-150 meV below the energy of atomic IS excitation
but still above the bottom of the estimated IS excitonic dispersion.

Finally, we analyze the possibility of condensation of the HS bi-excitons. To this end, we view the HS state as a bound pair of two $T_{1g}$ IS states with different orbital characters. 
This leads to each of the three HS orbital states to have a sizeable hopping only along one of the cubic axes  with the amplitude roughly estimated by
\begin{equation}
t_{\text{HS-LS}}=\frac{\sqrt{2} t_{\text{IS-LS}}^2} {\epsilon_{\text{HS}}+\epsilon_{\text{LS}}-2\epsilon_{\text{IS}} }.
\end{equation}
The upper bound for $t_{\text{HS-LS}}=\tfrac{1}{2\sqrt{2}}t_{\text{IS-LS}}$ is obtained assuming $\epsilon_{\text{HS}}=\epsilon_{\text{IS}}$ and $\epsilon_{\text{IS}}-\epsilon_{\text{LS}}=4t_{\text{IS-LS}}$ (dictated by the stability of the normal state in the absence of the external field).
This yields the bandwidth of the HS dispersion a factor of $4\sqrt{2}$ smaller than that of the IS dispersion. A more realistic assumption $\epsilon_{\text{HS}}\approx\epsilon_{\text{LS}}$ yields an order of magnitude
difference between the IS and HS bandwidths. Should the field-induced transition involve HS bi-excitons, their low mobility makes EC an unlikely competitor with SSO driven by
HS-HS repulsion of the order of 70 meV~\cite{zhang12}.

Therefore, the present results suggest the picture of \laco* as a gas of mobile IS excitons on the background of LS ground states interacting via attractive interaction, which leads to formation of immobile bi-excitations, the HS states. The high mobility of the IS excitons allows the lowest excitation of the system to be a wave-like IS state despite the likely $\epsilon_{\text{LS}}<\epsilon_{\text{HS}}<\epsilon_{\text{IS}}$ order of the 
atomic excitation energies. Wave-like character of the low-energy excitations may explain the absence of spin-state (and Co-O bond-length) disproportionation in the intermediate temperature 
range. The fact that these excitations do not carry charge is consistent with the insulating behavior in this regime. As the thermally induced concentration of IS excitations grows, bi-excitations, HS states, are formed. 
The on-set of the bad metallic state shall be viewed of melting of the excitons into free electrons and holes. Quantitative investigation of the present scenario is beyond the scope of this paper.
We are not aware of any studies that take the high mobility of the IS excitations into account. In particular, DMFT calculations~\cite{krapek12,zhang12} do not include the IS mobility in the normal state, since this involves two-fermion inter-site correlations. Nevertheless, DMFT can capture the field-induced exciton condensation when the correlations become static.  

{\bf Conclusions.}
We have used two-orbital Hubbard model to simulate the effect of an external magnetic field on the ordering transition in the vicinity of spin-state crossover. We find that the 
$dh_c/dT>0$ slope observed in \laco* is consistent with exciton condensation, but inconsistent with SSO. We show that the field-induced transition is of the insulator-to-insulator type.
We have estimated dispersion of the IS and HS excitations of the LS ground state and found sizeable bandwidths for the IS excitations of the order of several 100 meV, while the bandwidth
of the HS excitations is an order of magnitude smaller. We conclude that the field-induced transition is a Bose-Einstein condensation of the IS excitons. The mobility of IS excitations is a key property of the low-temperature
regime that has to be taken into account in description of \laco*.

\begin{acknowledgments}
The authors thank M.~Haverkort, A.~Hariki, V.~Pokorn\'{y}, and Z.~Jir\'ak for fruitful discussions.
This work has received funding from the European Research Council (ERC) under the European Union's Horizon 2020 research and innovation programme (grant agreement No. 646807-EXMAG).
\end{acknowledgments}



\bibliography{ECfield}

\end{document}